\documentclass{webofc}
\usepackage[varg]{txfonts}   
\usepackage{braket}
\usepackage{slashed}
\newcommand{\sm}{\text{SM}}
\newcommand{\np}{\text{NP}}
\newcommand{\hqe}{\text{HQE}}
\newcommand{\order}{\mathcal{O}}
\newcommand{\oper}{\mathcal{O}}
\newcommand{\wrt}{\text{w.r.t.}}
\newcommand{\lfu}{\text{LFU}}
\newcommand{\qcd}{\text{QCD}}
\newcommand{\lhc}{\text{LHC}}
\newcommand{\hami}{\mathcal{H}}
\newcommand{\eff}{\text{eff}}
\newcommand{\hc}{\text{h.c.}}
\newcommand{\ckm}{\text{CKM}}
\newcommand{\diff}{\text{d}}
\newcommand{\re}{\text{Re}}
\newcommand{\im}{\text{Im}}
\newcommand{\tprod}{\mathbb{T}}
\newcommand{\eom}{\text{EoM}}
\newcommand{\Dslash}{\slashed{D}}
\newcommand{\vslash}{\slashed{v}}
\newcommand{\matrice}{\mathcal{M}}
\newcommand{\mupi}{\hat{\mu}_\pi^2}
\newcommand{\muG}{\hat{\mu}_G^2}
\newcommand{\rhoD}{\hat{\rho}_D^3}
\newcommand{\rhoLS}{\hat{\rho}_{LS}^3}
\newcommand{\ope}{\text{OPE}}
\newcommand{\eq}{\text{Eq.}}
\newcommand{\gev}{\text{GeV}}
\newcommand{\br}{\mathcal{B}}
\newcommand{\bp}{\text{BP}}
\newcommand{\nlo}{\text{NLO}}
\begin{document}
\title{Inclusive $b \to \set{u, c} \, \ell^- \, \bar{\nu}_\ell$ modes: polarized/unpolarized $\Lambda_b$}
%
%

\author{\firstname{Francesco} \lastname{Loparco}\inst{1}\fnsep\thanks{\email{francesco.loparco1@ba.infn.it}}}

\institute{Istituto Nazionale di Fisica Nucleare, Sezione di Bari, Via Orabona 4, I-70126 Bari, Italy}

\abstract{%
The increasing number of flavour anomalies motivates the investigation of new processes where tensions similar to the observed ones may emerge.
It is necessary to identify observables sensitive to physics beyond the Standard Model.
The analysis which follows concerns the inclusive semileptonic decays of polarized beauty baryons, computed through the Heavy Quark Expansion at $\order(1 / m_b^3)$ and at the leading order in $\alpha_s$.
New Physics interactions have been taken into account, extending the Standard Model $b \to U \, \ell \, \bar{\nu}_\ell$ low-energy Hamiltonian, where $U = \set{c, u}$ and $\ell = \set{e, \mu, \tau}$, including the full set of $D = 6$ operators with left-handed neutrinos.
Among the possible observables one can consider, the ones depending on the spin of the decaying baryon are very appealing and can be considered for physics programmes of future facilities, such as FCC-ee.
}
\maketitle
\section{Introduction}
\label{intro}
Deviations in a number of observables \wrt\ the Standard Model (\sm) predictions, the \emph{flavour anomalies}, have been recently detected.
They represent a motivation for searching signals of New Physics (\np).
Anomalies have been observed in tree-level and loop-induced decays of $B$, $B_s$ and $B_c$ mesons \cite{Fajfer:2012vx, Alguero:2021anc}.
Particularly interesting, hints of lepton flavour universality (\lfu) violations have been collected.
It is necessary to investigate other heavy hadron decay processes to get a full comprehension of \lfu, for this reason inclusive semileptonic beauty baryon decays have been analyzed \cite{Colangelo:2020vhu} treating the nonperturbative effects of strong interactions in a systematic way, exploiting an
expansion in the inverse heavy quark mass \cite{Bigi:1993fe,Chay:1990da}.
The calculation of the inclusive semileptonic decay width $H_b \to X_{c, u} \, \ell^- \, \bar{\nu}_\ell$ of a polarized heavy hadron $H_b$ will be described below, expanding the hadronic matrix elements at $\order(1 / m_b^3)$ in the heavy quark expansion (\hqe), at leading order in $\alpha_s$.
Observables have been computed for the transition $\Lambda_b \to X_{c, u} \, \ell^- \, \bar{\nu}_\ell$ for non vanishing charged lepton mass and considering the \np\ generalization of the \sm\ in the effective Hamiltonian, including all the $D = 6$ semileptonic operators with left-handed neutrinos \cite{Biancofiore:2013ki, Colangelo:2018cnj, Bhattacharya:2018kig, Colangelo:2016ymy, Mannel:2017jfk, Kamali:2018fhr, Kamali:2018bdp}.
A recent improvement has been achieved in \cite{Moreno:2022goo} where $\alpha_s$ corrections has been computed for the inclusive $B \to X_c \, \tau \, \bar{\nu}_\tau$ decay width and leptonic invariant mass spectrum up to $\order(1 / m_b^3)$ in the \hqe.

At \lhc\ the $\Lambda_b$ is produced unpolarized \cite{Aaij:2013hzx, Aad:2014swk, Sirunyan:2018wjk, Aaij:2020iux} because the $b$ quark is produced through strong interactions, a sizeable longitudinal $\Lambda_b$ polarization is expected for $b$ quarks produced in $Z$ and $t$ quark decays \cite{Buskulic:1995aqx, Abbiendi:1998wmk, Abreu:1999hkl}.

\section{Generalized effective Hamiltonian}
\label{sec-1}

Let us consider a beauty hadron $H_b$ with spin $s$.
The inclusive semileptonic decays $H_b(p, s) \to X_U(p_X) \, \ell^-(p_\ell) \, \bar{\nu}_\ell(p_\nu)$, induced by the quark transition $b \to U$, with $U = \set{c, u}$, are described by the general low-energy Hamiltonian which extends the \sm\ one:
\begin{equation}
\label{effective_hamiltonian}
\hami_\eff^{b \to U \, \ell \, \bar{\nu}} = \frac{G_F \, V_{Ub}}{\sqrt{2}} \, \left[ ( 1 + \epsilon_V^\ell ) \, \oper_\sm + \epsilon_S^\ell \, \oper_S + \epsilon_P^\ell \, \oper_P + \epsilon_T^\ell \, \oper_T + \epsilon_R^\ell \, \oper_R \right] + \hc \;,
\end{equation}
that comprises the Fermi constant $G_F$, the \ckm\ matrix element $V_{Ub}$ and the full set of $D = 6$ semileptonic $b \to U$ operators with left-handed neutrinos: $\oper_{\sm,R} = \big[ \bar{u} \, \gamma_\mu \, (1 \mp \gamma_5) \, b \big] \, \big[ \bar{\ell} \, \gamma^\mu \, (1 - \gamma_5) \, \nu_\ell \big]$, $\oper_S = \big[ \bar{u} \, b \big] \, \big[ \bar{\ell} \, (1 - \gamma_5) \nu_\ell \big]$, $\oper_P = \big[ \bar{u} \, \gamma_5 \, b \big] \, \big[ \bar{\ell} \, (1 - \gamma_5) \nu_\ell \big]$ and $\oper_T = \big[ \bar{u} \, \sigma_{\mu\nu} \, (1 - \gamma_5) \, b \big] \, \big[ \bar{\ell} \, \sigma^{\mu\nu} \, (1 - \gamma_5) \, \nu_\ell \big]$.
$\epsilon_{V, S, P, T, R}^\ell$ are complex lepton-flavour dependent couplings.
In the case of $\epsilon_i^\ell = 0$ one recovers the \sm\ case.
We keep $m_\ell \neq 0$ for all leptons $\ell = \set{e, \mu, \tau}$.

\section{Inclusive decay width}
\label{sec-2}

\begin{figure}[h]
\centering
\sidecaption
\includegraphics[width=5cm,clip]{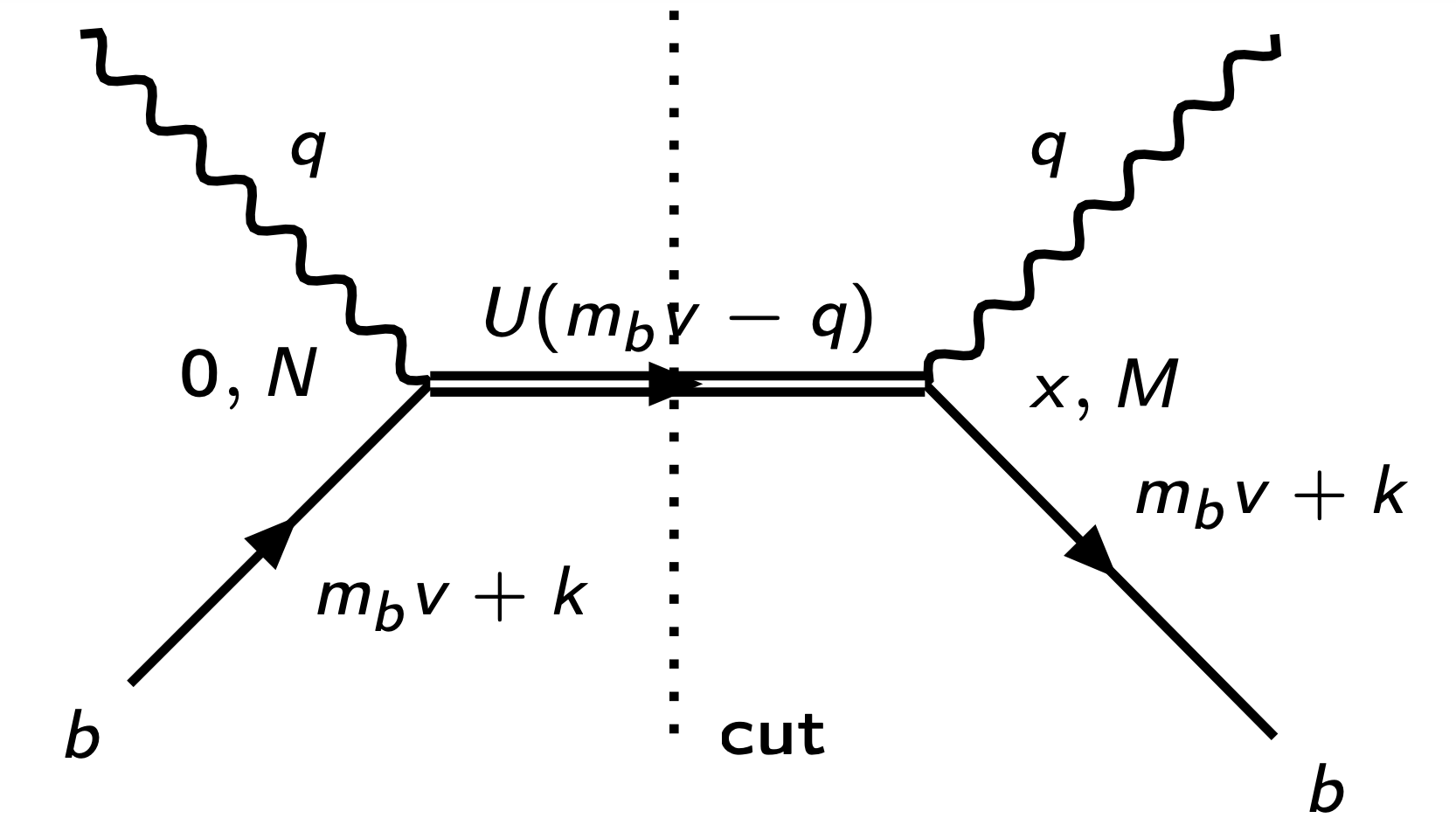}
\caption{Discontinuity of the forward amplitude across the cut of the semileptonic process.}
\label{discontinuity}
\end{figure}

Any $D = 6$ operator in \eq\ \eqref{effective_hamiltonian} can be written as the product of two currents, $\oper = J_M^{(i)} \, L^{M(i)}$, where $J_M^{(i)}$ and $L^{M(i)}$ are respectively the hadronic and the leptonic currents.
$M$ generically denotes the Lorentz indices of the currents.
The $H_b$ inclusive semileptonic differential decay width can be written as
\begin{equation}
\diff \Gamma = \underbrace{\diff \Sigma}_{\text{phace space}} \, \frac{G_F^2 \, |V_{Ub}|^2}{4 \, M_H} \, \sum_{i,j} \, g_i^* \, g_j \, \underbrace{(W^{ij})_{MN}}_{\text{hadronic tensor}} \, \underbrace{(L^{ij})^{MN}}_{\text{leptonic tensor}} \;,
\end{equation}
with the phase-space element $d \Sigma = (2 \, \pi)^4 \, d^4 q \, \delta^4(q - p_\ell - p_\nu) \, [d p_\ell] \, [d p_\nu]$.
The notation $[d p] = \frac{d^3 p}{(2 \, \pi)^3 \, 2 \, p^0}$ has been used.
$g_V = 1 + \epsilon_V^\ell$ and $g_{S, P, T, R} = \epsilon_{S, P, T, R}^\ell$ are the Wilson coefficients.\\
Using the optical theorem, the hadronic tensor $(W^{ij})_{MN} = \frac{1}{\pi} \, \im \left[ (T^{ij})_{MN} \right]$ is given in terms of the forward amplitude depicted in figure~\ref{discontinuity}
\begin{equation}
\label{Tdef}
(T^{ij})_{MN} = i \, \int \, d^4 x \, e^{- i \, q \cdot x} \, \braket{H_b(p, s) | \tprod \, \Big\{ {J_M^{(i)}}^\dagger(x) \, J_N^{(j)}(0) \Big\} | H_b(p, s)} \;.
\end{equation}
The hadron momentum $p = m_b \, v + k$ is written in terms of the 4-velocity $v$, the heavy quark mass $m_b$ and the residual momentum $k$, with $|k| \sim \order(\Lambda_\qcd)$.
The \qcd\ quark field $b(x)$ is redefined as $b(x) = e^{- i \, m_b \, v \cdot x} \, b_v(x)$, with $b_v(x)$ still \qcd\ quark field satisfying the \eom\ $b_v(x) = \left( P_+ + \frac{i \, \Dslash}{2 \, m_b} \right) \, b_v(x)$, where $P_\pm = \frac{1 \pm \vslash}{2}$ are the velocity projectors.

Introducing $p_X = m_b \, v + k - q$, the \eq\ \eqref{Tdef} becomes
\begin{equation}
(T^{ij})_{MN} = \braket{H_b(v, s) | \bar{b}_v(0) \, \overline{\Gamma}_M^{(i)} \, \frac{1}{\slashed{p}_X - m_U} \, \Gamma_N^{(j)} \, b_v(0) | H_b(v, s)} \;,
\end{equation}
and considering $|k| \sim \order(\Lambda_\qcd)$ we have
\begin{equation}
(T^{ij})_{MN} = \sum_{n=0}^{+ \infty} \, \braket{H_b(v, s) | \bar{b}_v(0) \, \overline{\Gamma}_M^{(i)} \, (\slashed{p}_U + m_U) \, ( i \, \Dslash \, (\slashed{p}_U + m_U))^n \, \Gamma_N^{(j)} \, b_v(0) | H_b(v, s)} \, \frac{(-1)^n}{\Delta_0^{n+1}} \;,
\end{equation}
where $\overline{\Gamma}_M^{(i)} = \gamma^0 \, \Gamma_M^{(i)\dagger} \, \gamma^0$, $p_U = p_X - k$ and $\Delta_0 = p_U^2 - m_U^2$.

Using the trace formalism \cite{Dassinger:2006md}, we can write the n-th term in the series as
\begin{equation}
\label{trace_formalism}
\begin{split}
& \braket{H_b(v, s) | \bar{b}_v(0) \, \overline{\Gamma}_M^{(i)} \, (\slashed{p}_U + m_U) \, \underbrace{i \, \Dslash \, (\slashed{p}_U + m_U) \dots i \, \Dslash \, (\slashed{p}_U + m_U)}_{\text{n times}} \, \Gamma_N^{(j)} \, b_v(0) | H_b(v, s)} = \\
& = \bigg[ \overline{\Gamma}_M^{(i)} \, ( \slashed{p}_U + m_U ) \, \prod_{k=1}^n \, \left[ \gamma_{\mu_k} \, (\slashed{p}_U + m_U) \right] \, \Gamma_N^{(j)} \bigg]_{ab} \, \underbrace{\braket{H_b(v, s) | \bar{b}_v(0) \, i D^{\mu_1} \, \dots \, i D^{\mu_n} \, b_v(0) | H_b(v, s)}_{ba}}_{(\matrice^{\mu_1 \dots \mu_n})_{ba}} \;.
\end{split}
\end{equation}
The expression in \eq\ \eqref{trace_formalism} involves the hadronic matrix elements
$(\matrice^{\mu_1 \dots \mu_n})_{ba}$ where $a$ and $b$ are Dirac indices.
These matrix elements can be written in terms of nonperturbative parameters, the number of which increases with the order of the expansion:
\begin{equation}
\order(1 / m_b^n) \dots
\begin{cases}
\order(1 / m_b^3)
\begin{cases}
\order(1 / m_b^2)
\begin{cases}
- 2 \, M_H \, \mupi = \braket{H_b|\bar{b}_v \, i D^\mu \, i D_\mu \, b_v|H_b} \\
2 \, M_H \, \muG = \braket{H_b|\bar{b}_v \, (- i \sigma_{\mu\nu}) \, i D^\mu \, i D^\nu \, b_v|H_b}
\end{cases} \\
2 \, M_H \, \rhoD = \braket{H_b|\bar{b}_v \, i D^\mu \, (i v \cdot D) \, i D_\mu \, b_v|H_b} \\
2 \, M_H \, \rhoLS = \braket{H_b|\bar{b}_v \, (- i \sigma_{\mu\nu}) \, i D^\mu \, (i v \cdot D) \, i D^\nu \, b_v|H_b}
\end{cases} \\
\dots
\end{cases} \;.
\end{equation}
To compute $\matrice^{\mu_1 \dots \mu_n}$ one can follow the methods proposed in \cite{Dassinger:2006md, Mannel:2010wj}.\\
In the case of baryons, the dependence on the spin $s_\mu$ in \eq\ \eqref{trace_formalism} must be kept into account as done up to $\order(1 / m_b^{-2})$ in \cite{Manohar:1993qn}.
In \cite{Colangelo:2020vhu} $\matrice^{\mu_1 \dots \mu_n}$ have been derived at $\order(1 / m_b^{-3})$ for a polarized baryon, extending the previous results obtained in \cite{Mannel:2017jfk, Kamali:2018bdp, Grossman:1994ax, Manohar:1993qn, Balk:1997fg}.

From the expressions of the matrix elements $\matrice^{\mu_1 \dots \mu_n}$ the hadronic tensor can be computed and expanded in Lorentz structures which depend on $v$, $q$ and $s$.
The results for the effective Hamiltonian \eq\ \eqref{effective_hamiltonian} are collected in \cite{Colangelo:2020vhu}.
The four-fold differential decay rate for the $H_b(p, s) \to X_U(p_X) \, \ell^-(p_\ell) \, \bar{\nu}_\ell(p_\nu)$ transition reads
\begin{equation}
\label{fully_diff_distr}
\frac{\diff^4 \Gamma}{\diff E_\ell \, \diff q^2 \, \diff q_0 \, \diff \cos \theta_P} = \frac{G_F^2 \, |V_{Ub}|^2}{32 \, (2 \, \pi)^3 \, M_H} \, \sum_{i, j} \, g_i^* \, g_j \, \frac{1}{\pi} \, \im \left[ (T^{ij})_{MN} \right] \, (L^{ij})^{MN} \;,
\end{equation}
where $p_\ell = (E_\ell, \vec{p}_\ell)$, $q_0 = q \cdot v$ and $\theta_P$ is the angle between the two 3-vectors $\vec{p}_\ell$ and $\vec{s}$ in the $H_b$ rest frame.
Double and single distributions are obtained integrating \eq\ \eqref{fully_diff_distr} over the phase-space \cite{Jezabek:1996ia}.
The full decay width can be obtained by performing all integrations and can be written as:
\begin{equation}
\Gamma(H_b \to X_U \, \ell^- \, \bar{\nu}_\ell) = \Gamma_b \, \sum_{i,j} \, g_i^* \, g_j \,  \, \left[ \mathcal{C}_{0}^{(i,j)} + \frac{\mupi}{m_b^2} \, \mathcal{C}_{\mupi}^{(i,j)} + \frac{\muG}{m_b^2} \, \mathcal{C}_{\muG}^{(i,j)} + \frac{\rhoD}{m_b^3} \, \mathcal{C}_{\rhoD}^{(i,j)} + \frac{\rhoLS}{m_b^3} \, \mathcal{C}_{\rhoLS}^{(i,j)} \right] \;,
\end{equation}
where $\Gamma_b = \frac{G_F^2 \, |V_{Ub}|^2 \, m_b^5}{192 \, \pi^3}$ is the \emph{partonic term} corresponding to a free quark decay.
The indices $i$ and $j$ runs over the contributions of the various operators and of their interferences.
The coefficients $\mathcal{C}^{(i,j)}$ can be found in \cite{Colangelo:2020vhu}.
The \ope\ breaks down in the endpoint region of the spectra, where singularities appear and which should be resummed in a $H_b$ shape function, the convolution with such a function smears the spectra at the endpoint. 
The effects of the baryon shape function has not been considered here.
Perturbative \qcd\ corrections have also been not included: in the \sm\ case they can be found in \cite{Czarnecki:1994bn, Jezabek:1996db, DeFazio:1999ptt, Trott:2004xc, Aquila:2005hq, Alberti:2014yda}.
\nlo\ \qcd\ corrections have also been recently computed in \cite{Mannel:2021zzr, Moreno:2022goo}.

\section{Results for $\Lambda_b \to X_{c, u} \, \ell \, \bar{\nu}_\ell$}
\label{sec-3}

In \cite{Colangelo:2020vhu} several observables for the modes $\Lambda_b \to X_{c, u} \, \ell \, \bar{\nu}_\ell$ have been studied.
Here we consider a few of them.
Further ones, as well as the input parameters, can be found in \cite{Colangelo:2020vhu}.
For the couplings $\epsilon_i^\ell$ in \eq\ \eqref{effective_hamiltonian} we fix three \np\ benchmark points, set in \cite{Colangelo:2018cnj, Shi:2019gxi} for $U = c$, while for $U = u$ we use the ranges fixed in \cite{Colangelo:2019axi}.
Using $G_F = 1.16637(1) \times 10^{-5} \, \gev^{-2}$, $|V_{cb}| = 0.042$ and $|V_{ub}| = 0.0037$, together with $\tau_{\Lambda_b} = 1.471(9) \, \text{ps}$ \cite{Workman:2022ynf},
we compute the inclusive $\Lambda_b$ branching fraction for the two quark transitions and for final $\tau$ and $\mu$ lepton.
The results in the \sm, for the central values of the parameters and neglecting \qcd\ corrections, are in table~\ref{inclusive_br_for_Lambda_b}.
\begin{table}
\centering
\caption{Inclusive semileptonic $\Lambda_b$ branching fractions in \sm, obtained for the central values of the parameters.}
\label{inclusive_br_for_Lambda_b}
\begin{tabular}{crcr}
\hline
\noalign{\smallskip}
\multicolumn{2}{c}{$b \to c$} & \multicolumn{2}{c}{$b \to u$} \\
\noalign{\smallskip}
\hline
\noalign{\smallskip}
$\br(\Lambda_b \to X_c \, \mu \, \bar{\nu}_\mu)$ & $11.0 \times 10^{-2}$ & $\br(\Lambda_b \to X_u \, \mu \, \bar{\nu}_\mu)$ & $11.65 \times 10^{-4}$ \\
\noalign{\smallskip}
\hline
\noalign{\smallskip}
$\br(\Lambda_b \to X_c \, \tau \, \bar{\nu}_\tau)$ & $2.4 \times 10^{-2}$ & $\br(\Lambda_b \to X_u \, \tau \, \bar{\nu}_\tau)$ & $2.75 \times 10^{-4}$ \\
\noalign{\smallskip}
\hline
\end{tabular}
\end{table}
For comparison, the available measurements are $\br(\Lambda_b \to \Lambda_c \, \ell^- \, \bar{\nu}_\ell + \text{anything}) = ( 10.9 \pm 2.2 ) \times 10^{-2}$, with $\ell = e, \mu$ and $\br(\Lambda_b \to p \, \mu^- \, \bar{\nu}_\mu) = ( 4.1 \pm 1.0 ) \times 10^{-4}$\cite{Workman:2022ynf}.

\begin{figure}[h]
\centering
\includegraphics[width=5cm,clip]{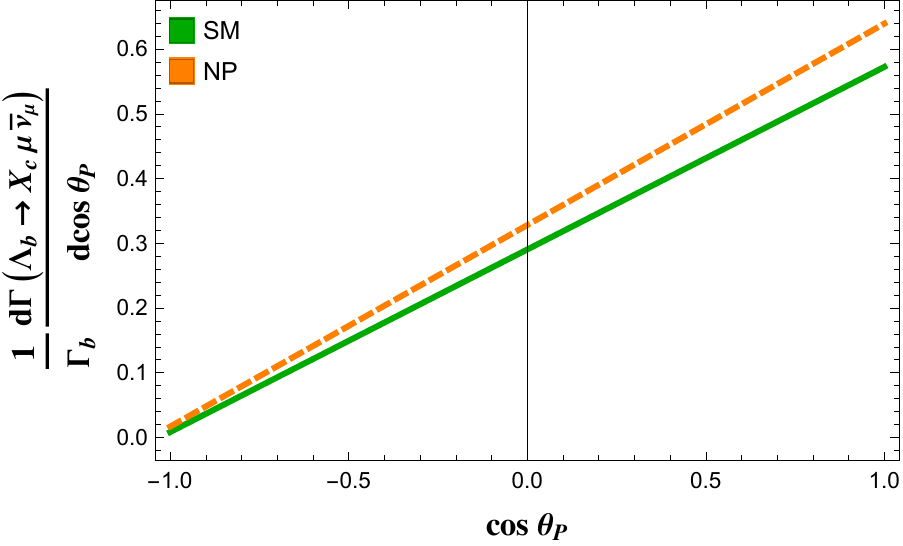} \hspace{0.5cm}
\includegraphics[width=5cm,clip]{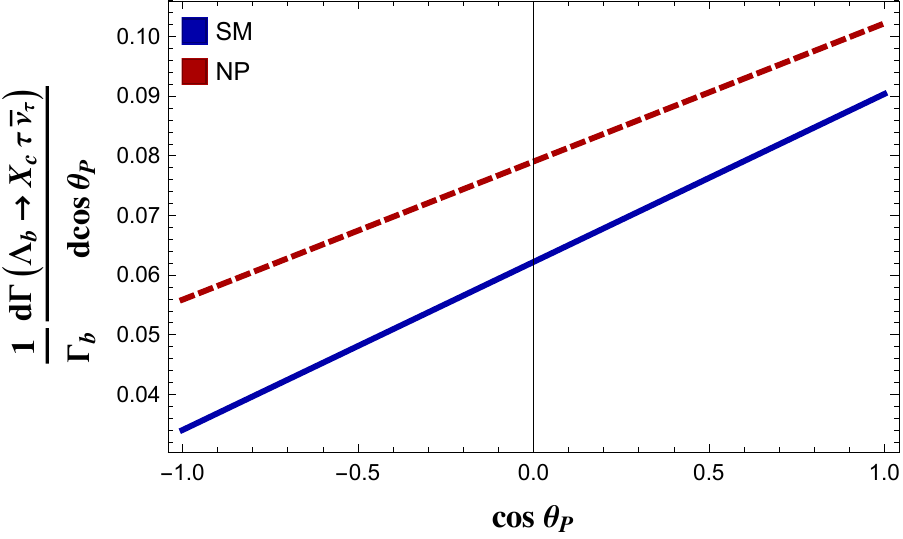}
\includegraphics[width=5cm,clip]{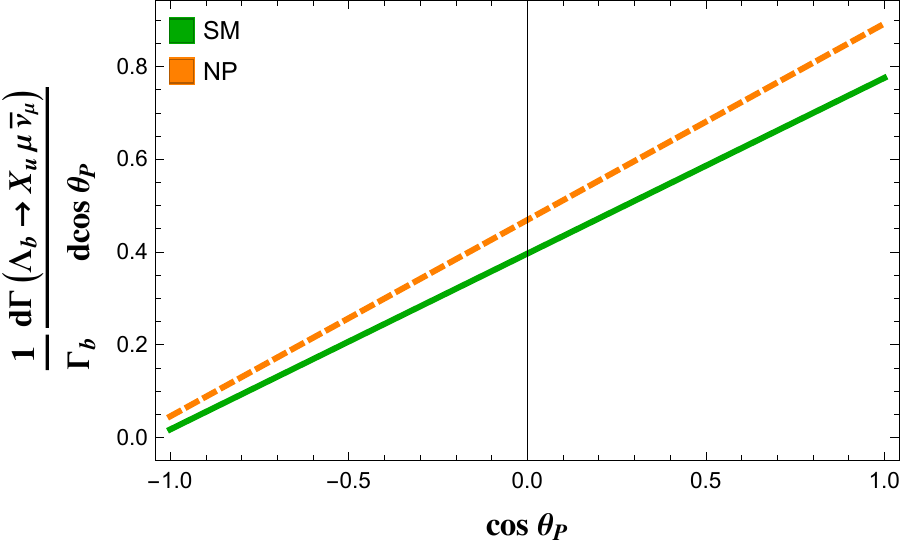} \hspace{0.5cm}
\includegraphics[width=5cm,clip]{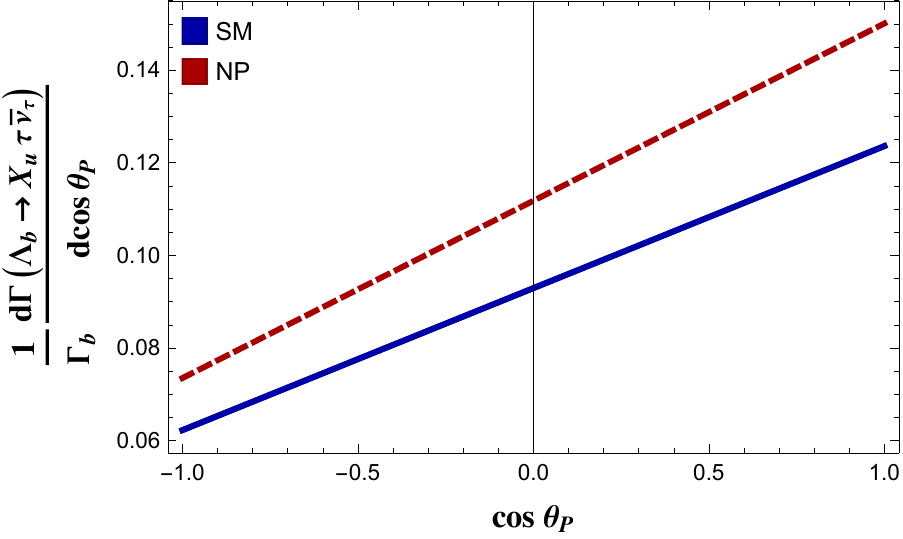}
\caption{$\frac{1}{\Gamma_b} \, \frac{\diff \Gamma}{\diff \cos \theta_P}$ distribution for $\Lambda_b \to X_U \, \ell \, \bar{\nu}_\ell$, with $U = c$ (top row), $U = u$ (bottom row), $\ell = \mu$ (left column) and $\ell = \tau$ (right column).
The solid line is the \sm\ result, the dashed line the \np\ result at the benchmark point.}
\label{linear}
\end{figure}

\begin{figure}[h]
\centering
\sidecaption
\includegraphics[width=5cm,clip]{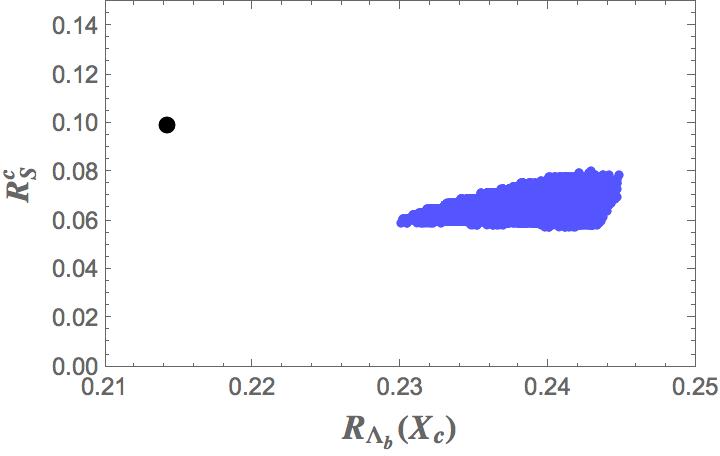}
\caption{Correlation between $R_{\Lambda_b}(X_c)$ and the ratio $R_S^c$ of the slopes of the $\frac{\diff \Gamma}{\diff \cos \theta_P}$ distribution.
The dot corresponds to \sm, the broad region to \np\ with the effective couplings varied as specified in the text.}
\label{corr}
\end{figure}

For inclusive semileptonic $\Lambda_b$ decays it is interesting to consider a ratio analogous to $R{(D^*)}$ for $B$ meson, to compare the $\tau$ and the muon mode using a quantity in which several theoretical uncertainties cancel:
\begin{equation}
\label{ratio}
R_{\Lambda_b}(X_U) = \frac{\Gamma(\Lambda_b \to X_U \, \tau \, \bar{\nu}_\tau)}{\Gamma(\Lambda_b \to X_U \, \mu \, \bar{\nu}_\mu)} \qquad \text{with} \qquad U = u, c \;.
\end{equation}
Another ratio sensitive to \lfu\ violating \np\ effects can be defined.
It can be constructed from the distribution $\frac{\diff \Gamma(\Lambda_b \to X_U \, \ell \, \bar{\nu}_\ell)}{\diff \cos \theta_P} = A_\ell^U + B_\ell^U \, \cos \theta_P$.
The dependence of $\frac{\diff \Gamma(\Lambda_b \to X_U \, \ell \, \bar{\nu}_\ell)}{\diff \cos \theta_P}$ on $\cos \theta_P$ is linear, and \np\ contributions modify both the slope and the intercept of the curve as displayed in figure \ref{linear}.
Similarly to the ratio defined in \eq\ \eqref{ratio}, we define the ratio of the slopes $R_S^U = B_\tau^U / B_\mu^U$ which has a definite value in \sm, and can deviate due to \np\ as specified in table~\ref{ratio_R_Lambda_b}.
\begin{table}[h]
\centering
\caption{Ratios $R_{\Lambda_b}(X_u)$ and $R_{\Lambda_b}(X_c)$ (left) and ratios $R_S^u$ and $R_S^c$ (right) for \sm\ and \np\ at the \bp.}
\label{ratio_R_Lambda_b}
\begin{tabular}{ccccc}
\hline
\noalign{\smallskip}
& $R_{\Lambda_b}(X_u)$ & $R_{\Lambda_b}(X_c)$ & $R_S^u$ & $R_S^c$ \\
\noalign{\smallskip}
\hline
\noalign{\smallskip}
\sm\ & $0.234$ & $0.214$ & $0.081$ & $0.100$ \\
\noalign{\smallskip}
\hline
\noalign{\smallskip}
\np\ & $0.238$ & $0.240$ & $0.091$ & $0.074$ \\
\noalign{\smallskip}
\hline
\end{tabular}
\end{table}
A correlation between $R_{\Lambda_b}(X_U)$ and $R_S^U$ can be constructed.
As an example, for $\Lambda_b \to X_c \, \tau \, \bar{\nu}_\tau$ with the effective Hamiltonian extended including a tensor operator, we vary the couplings $\big( \re[\epsilon_T^\mu], \im[\epsilon_T^\mu] \big)$ and $\big( \re[\epsilon_T^\tau], \im[\epsilon_T^\tau] \big)$ in the regions determined in \cite{Colangelo:2018cnj}.
The correlation plot in figure~\ref{corr} shows that the (challenging) measurement of the two ratios would discriminate \sm\ and \np.

\section{Conclusions}

The calculation of the fully differential inclusive semileptonic decay width of a polarized heavy hadron at $\order(1 / m_b^3)$ in the \hqe, at leading order in $\alpha_s$ and for non vanishing charged lepton mass has been described.
\np\ generalization of the \sm\ has been considered in the effective Hamiltonian in \eq \eqref{effective_hamiltonian} including all the $D = 6$ semileptonic operators with left-handed neutrinos.
The correlation plot in figure~\ref{corr} shows that the ratios $R_{\Lambda_b}(X_c)$ and $R_S^c$ can assume different values in the \sm\ and when \np\ is included.
Although this measurement is experimentally challenging, it represents a suitable observable to test violation of \lfu.

\section*{Acknowledgments}

I thank P. Colangelo and F. De Fazio for collaboration.
This study has been carried out within the INFN project (Iniziativa Specifica) QFT-HEP.

\nocite{}

\bibliographystyle{woc}
\bibliography{biblio}

\begin{thebibliography}{36}

\bibitem{Fajfer:2012vx}
S.~Fajfer, J.F. Kamenik, I.~Nisandzic, Phys. Rev. \textbf{D85}, 094025 (2012),
  \texttt{1203.2654}

\bibitem{Alguero:2021anc}
M.~Alguer\'o, B.~Capdevila, S.~Descotes-Genon, J.~Matias, M.~Novoa-Brunet,
  \emph{{$\boldsymbol{b\to s\ell\ell}$ global fits after Moriond 2021
  results}}, in \emph{{55th Rencontres de Moriond on QCD and High Energy
  Interactions}} (2021), \texttt{2104.08921}

\bibitem{Colangelo:2020vhu}
P.~Colangelo, F.~De~Fazio, F.~Loparco, JHEP \textbf{11}, 032 (2020),
  \texttt{2006.13759}

\bibitem{Bigi:1993fe}
I.I.Y. Bigi, M.A. Shifman, N.G. Uraltsev, A.I. Vainshtein, Phys. Rev. Lett.
  \textbf{71}, 496 (1993), \texttt{hep-ph/9304225}

\bibitem{Chay:1990da}
J.~Chay, H.~Georgi, B.~Grinstein, Phys. Lett. B \textbf{247}, 399 (1990)

\bibitem{Biancofiore:2013ki}
P.~Biancofiore, P.~Colangelo, F.~De~Fazio, Phys. Rev. \textbf{D87}, 074010
  (2013), \texttt{1302.1042}

\bibitem{Colangelo:2018cnj}
P.~Colangelo, F.~De~Fazio, JHEP \textbf{06}, 082 (2018), \texttt{1801.10468}

\bibitem{Bhattacharya:2018kig}
S.~Bhattacharya, S.~Nandi, S.~Kumar~Patra, Eur. Phys. J. C \textbf{79}, 268
  (2019), \texttt{1805.08222}

\bibitem{Colangelo:2016ymy}
P.~Colangelo, F.~De~Fazio, Phys. Rev. \textbf{D95}, 011701 (2017),
  \texttt{1611.07387}

\bibitem{Mannel:2017jfk}
T.~Mannel, A.V. Rusov, F.~Shahriaran, Nucl. Phys. B \textbf{921}, 211 (2017),
  \texttt{1702.01089}

\bibitem{Kamali:2018fhr}
S.~Kamali, A.~Rashed, A.~Datta, Phys. Rev. D \textbf{97}, 095034 (2018),
  \texttt{1801.08259}

\bibitem{Kamali:2018bdp}
S.~Kamali, Int. J. Mod. Phys. A \textbf{34}, 1950036 (2019),
  \texttt{1811.07393}

\bibitem{Moreno:2022goo}
D.~Moreno (2022), \texttt{2207.14245}

\bibitem{Aaij:2013hzx}
R.~Aaij et~al. (LHCb), Phys. Lett. B \textbf{724}, 27 (2013),
  \texttt{1302.5578}

\bibitem{Aad:2014swk}
G.~Aad et~al. (ATLAS), Phys. Rev. D \textbf{89}, 092009 (2014),
  \texttt{1404.1071}

\bibitem{Sirunyan:2018wjk}
A.M. Sirunyan et~al. (CMS), Phys. Rev. D \textbf{97}, 072010 (2018),
  \texttt{1802.04867}

\bibitem{Aaij:2020iux}
R.~Aaij et~al. (LHCb), JHEP \textbf{06}, 110 (2020), \texttt{2004.10563}

\bibitem{Buskulic:1995aqx}
D.~Buskulic et~al. (ALEPH), Phys. Lett. B \textbf{365}, 437 (1996)

\bibitem{Abbiendi:1998wmk}
G.~Abbiendi et~al. (OPAL), Phys. Lett. B \textbf{444}, 539 (1998),
  \texttt{hep-ex/9808006}

\bibitem{Abreu:1999hkl}
P.~Abreu et~al. (DELPHI), Phys. Lett. B \textbf{474}, 205 (2000)

\bibitem{Dassinger:2006md}
B.M. Dassinger, T.~Mannel, S.~Turczyk, JHEP \textbf{03}, 087 (2007),
  \texttt{hep-ph/0611168}

\bibitem{Mannel:2010wj}
T.~Mannel, S.~Turczyk, N.~Uraltsev, JHEP \textbf{11}, 109 (2010),
  \texttt{1009.4622}

\bibitem{Manohar:1993qn}
A.V. Manohar, M.B. Wise, Phys. Rev. D \textbf{49}, 1310 (1994),
  \texttt{hep-ph/9308246}

\bibitem{Grossman:1994ax}
Y.~Grossman, Z.~Ligeti, Phys. Lett. B \textbf{332}, 373 (1994),
  \texttt{hep-ph/9403376}

\bibitem{Balk:1997fg}
S.~Balk, J.G. Korner, D.~Pirjol, Eur. Phys. J. C \textbf{1}, 221 (1998),
  \texttt{hep-ph/9703344}

\bibitem{Jezabek:1996ia}
M.~Jezabek, L.~Motyka, Acta Phys. Polon. B \textbf{27}, 3603 (1996),
  \texttt{hep-ph/9609352}

\bibitem{Czarnecki:1994bn}
A.~Czarnecki, M.~Jezabek, J.H. Kuhn, Phys. Lett. B \textbf{346}, 335 (1995),
  \texttt{hep-ph/9411282}

\bibitem{Jezabek:1996db}
M.~Jezabek, L.~Motyka, Nucl. Phys. B \textbf{501}, 207 (1997),
  \texttt{hep-ph/9701358}

\bibitem{DeFazio:1999ptt}
F.~De~Fazio, M.~Neubert, JHEP \textbf{06}, 017 (1999), \texttt{hep-ph/9905351}

\bibitem{Trott:2004xc}
M.~Trott, Phys. Rev. D \textbf{70}, 073003 (2004), \texttt{hep-ph/0402120}

\bibitem{Aquila:2005hq}
V.~Aquila, P.~Gambino, G.~Ridolfi, N.~Uraltsev, Nucl. Phys. B \textbf{719}, 77
  (2005), \texttt{hep-ph/0503083}

\bibitem{Alberti:2014yda}
A.~Alberti, P.~Gambino, K.J. Healey, S.~Nandi, Phys. Rev. Lett. \textbf{114},
  061802 (2015), \texttt{1411.6560}

\bibitem{Mannel:2021zzr}
T.~Mannel, D.~Moreno, A.A. Pivovarov, Phys. Rev. D \textbf{105}, 054033 (2022),
  \texttt{2112.03875}

\bibitem{Shi:2019gxi}
R.X. Shi, L.S. Geng, B.~Grinstein, S.~J\"ager, J.~Martin~Camalich, JHEP
  \textbf{12}, 065 (2019), \texttt{1905.08498}

\bibitem{Colangelo:2019axi}
P.~Colangelo, F.~De~Fazio, F.~Loparco, Phys. Rev. D \textbf{100}, 075037
  (2019), \texttt{1906.07068}

\bibitem{Workman:2022ynf}
R.L. Workman (Particle Data Group), PTEP \textbf{2022}, 083C01 (2022)

\end{thebibliography}

\end{document}